\documentclass[conference]{IEEEtran}
\IEEEoverridecommandlockouts
\usepackage{cite}
\usepackage{amsmath,amssymb,amsfonts}
\usepackage{algorithmic}
\usepackage{graphicx}
\usepackage[table]{xcolor}
\usepackage{rotating}
\usepackage{textcomp}
\usepackage{array}
\usepackage{booktabs}
\usepackage{multirow}
\usepackage{wasysym}
\usepackage{xcolor}
\def\BibTeX{{\rm B\kern-.05em{\sc i\kern-.025em b}\kern-.08em
    T\kern-.1667em\lower.7ex\hbox{E}\kern-.125emX}}

\usepackage{pgfplots}
\usepackage{pgfplotstable}

\usepackage{wrapfig}
\usepackage{lscape}
\usepackage{rotating}
\usepackage{epstopdf}
\usetikzlibrary{shapes,snakes}

\usepackage{pgf-pie,etoolbox}
\usepackage{bchart}
\pgfplotsset{compat=1.16}

\begin{document}

\title{Evaluating Dissemination and Implementation Strategies to Develop Clinical Software}

\author{\IEEEauthorblockN{Gastón Márquez}
\IEEEauthorblockA{\textit{Departamento de Informática} \\
\textit{Universidad Técnica Federico Santa María}\\
Valparaíso, Chile \\
gaston.marquez@usm.cl}
\and
\IEEEauthorblockN{Carla Taramasco}
\IEEEauthorblockA{\textit{Escuela de Ingeniería Informática} \\
\textit{Universidad de Valparaíso}\\
Valparaíso, Chile \\
carla.taramasco@uv.cl}
}

\maketitle

\begin{abstract}
Clinical software has become a significant contribution to support clinical management and intra-hospital processes.  
In this regard, the success or failure of clinical software is mostly yielded on a suitable requirements elicitation process.
Although several techniques and approaches address this process, the complexity of clinical services and the variety of clinicians involved in those services make it challenging to elicit requirements.
To address this concern, in our previous work, we have proposed the D\&I Framework. 
This collaborative technique translates clinical priorities into guidelines for eliciting software requirements in the healthcare context using implementation and dissemination strategies.
This article evaluates the functionalities and tasks implemented in a clinical bed management system whose requirements were elicited using the D\&I Framework.
We focused on evaluating clinicians' usability expectation levels using a specific questionnaire executed in 2018 and 2020.
The results show that, in comparison with the first release (2018) and the last one (2020), clinicians perceive an improvement in the functionalities and tasks implemented in the system.
This study introduces the effects of implementation and dissemination strategies to elicit pragmatic clinical requirements.

\end{abstract}


\section{Introduction}
\label{sec:introduction}

Software Engineering is a discipline that applies systematic approaches to developing, operating and maintaining software \cite{bourque2014}.
This discipline is made up of different phases that, together, allow the software to be developed effectively and efficiently. 
One of these phases corresponds to Requirements Engineering \cite{Van2009}. 
This phase aims at the elicitation, analysis, specification and validation of software requirements, which is managed throughout the  whole life cycle of the software product. 
It also expresses the needs and constraints that the software product must satisfy and define the basis for other software engineering phases, such as design, development, testing, maintenance, and processes.

Software products have been helpful in supporting management and productivity in several economic sectors, including healthcare.
Much of daily clinical work is supported by software that bridge several clinical processes from health care management to more specialized procedures, such as surgeries \cite{kaplan2009}. 
To develop clinical software, developers must often face several challenges.  
One of these challenges is to understand the real clinical problem that the software must address. 
Since clinical processes are complex and involve people with different perspectives, the success or failure of clinical software depends on the correct elicitation, identification, and characterization of requirements. 
Therefore, inadequate identification of clinical software requirements can lead to the rejection of the software and reduce clinicians' expectations \cite{heeks2006}.

To support clinical software developers, we have proposed the D\&I Framework \cite{marquez2020}.  
This collaborative technique identifies and characterizes clinical priorities that should be implemented in the software. 
These clinical priorities are mapped into software requirement elicitation guidelines through implementation and dissemination strategies \cite{brownson2017}. 
These strategies correspond to knowledge blocks obtained from evidence-based research that allow the distribution and integration of innovation towards a target audience (in our case, clinicians).

This article reports a study about the perception of clinicians regarding a bed management software called SIGICAM, whose software requirements were elicited and defined using the D\&I Framework. 
Since several alternatives and perspectives exist to evaluate a software product, in this study we focused on evaluating SIGICAM's functionalities and tasks based on the usability expectation levels of clinicians using the Health-ITUES questionnaire \cite{Yen2010}. 
We conducted the questionnaire in two instances in order to compare and analyze the clinicians' perceptions concerning the system's improvements. 
One instance was in November 2018 and the other one was in February 2020. 
Our contribution is an evaluative study regarding the support of implementation and dissemination strategies for software developers in order to elicit clinical software requirements.

This article is organized as follows:
Section \ref{sec:problem} describes the problem;
Section \ref{sec:relatedWork} details the related work;
Section \ref{sec:framework} introduces the D\&I Framework;
Section \ref{sec:caseStudy} describes the case study design;
Section \ref{sec:results} shows the results of the study;
Section \ref{sec:discussion} discuss the findings;
and Section \ref{sec:conclusions} concludes and describes the future work;
\section{Problem statement}
\label{sec:problem}

Clinical services are organized at different levels of complexity. 
These levels, in turn, are made up of different clinical processes that involve a range of activities that impact on patient care. 
These activities aim to achieve effective and efficient care for patients. 
In this regard, the role of clinical software in these processes is to support clinical management in order to optimize resources and procedures for providing quality care to patients \cite{Strasser2011}. 

Although the Software Engineering discipline has well-established techniques, methods, and processes for developing and implementing software in any domain, several factors strongly influence the success (or failure) of clinical software in the context of health. 
These factors can arise in all areas of software development, but they are more critical in the area of software requirements. 
According to Bourque \textit{et al.} \cite{bourque2014}, software requirements represent the needs and limitations of a software product that contribute to the solution of some real-world problem. 
This means that the software development team must fully understand the real problem that the software must address.

The software requirement community has proposed several techniques that allow requirements to be identified, described and characterized. 
Each of these techniques has a purpose (e.g., eliciting requirements, analyzing requirements, specifying requirements, among others); however, it is not clear how to use or combine such techniques to define software requirements in order to represent the real clinical problem.
The complexity of clinical services is not only related to clinical practice; it is also related to the diverse clinical and administrative professionals involved in the clinical processes \cite{Islam2016}. 
Moreover, clinical processes are composed of other internal sub-processes as well as professionals with different profiles. 
Some of these sub-processes correspond to strategic processes, suppliers, operational processes, laboratories, infrastructure, among others.
This implies that software development teams are forced to expand their scope of techniques to elicit requirements.

According to Tiwari \textit{et al.} \cite{tiwari2012}, an incorrect requirements elicitation process can directly affect the success of the software. 
Some studies (such as \cite{charette2005} \cite{doloi2013} \cite{cabana1999}) mention that poor requirements elicitation affects stakeholder expectations, leading to a rejection of software inclusion in clinical processes.
Moreover, other studies (such as \cite{johnson2011} \cite{chowdhury2007}) mention that if IT professionals do not have the skills for understanding the clinical problem, some issues may arise that compromise clinical software success. 
Some of these issues are as follows: (i) unclear project goal, (ii) poor planning and (iii) unrealistic scheduling or resource estimation \cite{johnson2006}.

In our previous work \cite{marquez2020}, we investigated implementation and dissemination sciences \cite{brownson2017} as an alternative to support clinical software developers to address the problem mentioned above.
Dissemination and implementation sciences use knowledge from evidence-based research to reduce barriers in order to deliver innovation and technology adoption in health care settings. 
On the one hand, dissemination science refers to the distribution of an intervention to a specific audience. 
On the other hand, implementation science focuses on the integration of new practices in a context. 
Both sciences provide a set of strategies that can be used in the context of software requirements to support developers in developing clinical software.
\section{Related Work}
\label{sec:relatedWork}

This section introduces studies that use implementation and dissemination sciences to promote the adoption of technology and software in health.

Vamos \textit{et al.} \cite{Vamos2020} discuss the importance of implementation science in informing the development of an eHealth application that enable providers to apply prenatal oral health guidelines during prenatal visits. 
In this study, the authors analyze which factors influence the implementation of a technology application in a clinical environment. 
The study results suggest that implementation science facilitates and drives guidelines that easily allow the eHealth application to translate interprofessional prenatal oral health guidelines into practice.

Richardson \textit{et al.} \cite{Richardson2012} address the development of Electronic Health Records (EHRs) in community settings. 
The authors propose an implementation science framework in informatics to qualitatively evaluate EHR implementations in communities. 
The authors provide an overview of the framework and specific methods for adapting the framework to the technology.
Similarly, the authors also discuss the contribution of the framework for health informatics research.

Glasgow \textit{et al.} \cite{Glasgow2014} propose key issues in eHealth from an implementation science perspective to illustrate how this science helps the adoption of technologies in health. 
The authors also use implementation science models to analyze the implication of these models in eHealth. 
More precisely, the authors provide examples of practical designs, measures and exemplary studies that address key issues in implementation science in eHealth.

Schoville \textit{et al.} \cite{Schoville2015} discuss how implementation and dissemination sciences help technology adoption in nursing practice. 
The authors emphasize that there is little knowledge of how health care facilities purchase, implement and adopt technology. 
In this regard, the technologies associated with implementation and dissemination sciences are key to successful adoption. 
Both sciences focus primarily on how end-users adopt technology through methods and interventions.

Dykes \textit{et al.} \cite{Dykes2020} apply the principles of implementation science to guide the deployment and use of the Hospital Clinical Surveillance (HCS) system.
Additionally, the authors develop a toolkit to promote optimal application, adoption, use and dissemination of the HCS. 
The authors describe that using the principles of implementation science is effective for learning about the effectiveness of the HCS and developing a toolkit to promote the system's main functionalities.

The studies discussed in this section confirm the relevance of implementation and dissemination sciences as a mechanism for translating clinical evidence-based practices into real-life contexts. 
Nevertheless, to the best of our knowledge, there is little discussion about how implementation and dissemination sciences support the elicitation of software requirements. 
Additionally, there is not enough evidence about comparative studies that discuss the clinicians' perceptions of clinical software developed and implemented using implementation and dissemination science strategies.
\section{The D\&I framework}
\label{sec:framework}

\begin{figure} [h]
  \centering
  \includegraphics[scale=0.2]{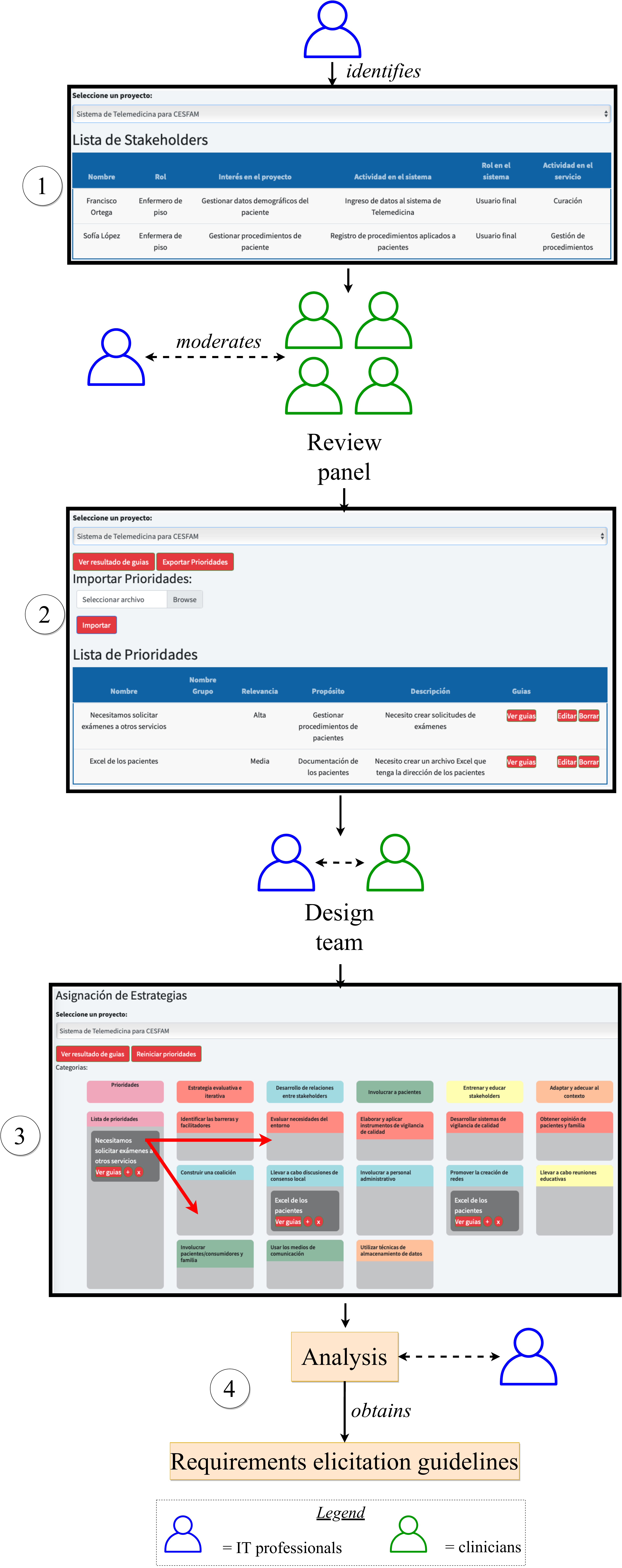}
  \caption{The D\&I Framework}
  \label{fig:Figures(web)}
\end{figure}

The D\&I Framework is a technique that suggests guidelines to elicit requirements in order to contextualize the clinical problem that the software must address (see Figure \ref{fig:Figures(web)}) \cite{marquez2020}. 
The D\&I Framework combines requirement elicitation techniques and clinical intervention-based implementation and dissemination strategies \cite{brownson2017} to support software development teams in implementing clinical software.
To suggest guidelines for requirements elicitation, the framework considers four stages. 
These stages are as follows: identification of project stakeholders ({\large \textcircled{\small 1}}), identification of clinical priorities ({\large \textcircled{\small 2}}), collaborative selection of implementation strategies ({\large \textcircled{\small 3}}), and analysis ({\large \textcircled{\small 4}}). 
In the following sections, we describe each stage in more detail.

\subsection{Identification of project stakeholders {\large \textcircled{\small 1}}}

This stage is focused on the identification and characterization of the project's main stakeholders. 
The project leader should identify the clinicians who influence the success (or failure) of the clinical software. 
Additionally, in this stage, the following information should be identified by each stakeholder:

\begin{itemize}
    \item \textit{Activity in the clinical service}: Each stakeholder has different responsibilities in the clinical service. Therefore, the project leader's task is to identify those stakeholders who have the main responsibilities in the clinical service in order to know their activities and daily work.
    
    \item \textit{Interest in the system}: Since the stakeholders are clinicians with different profiles and responsibilities, the project leader must know the stakeholders' interests and expectations in the clinical software.
    
    \item \textit{Role in the system}: The project leader must establish the different levels of responsibility that the stakeholders have in the clinical software.
\end{itemize}

Once the project leader identifies and registers the stakeholders in the D\&I Framework platform, a panel composed of the project's main stakeholders (including part of the development team) must be established.  
This panel has the mission of evaluating the clinical priorities that the clinical software should address.

\subsection{Identification of clinical priorities {\large \textcircled{\small 2}}}

This stage considers a workshop with all the stakeholders identified in the previous stage. 
In this workshop, all stakeholders should declare what clinical priorities the software should satisfy. 
It is important to note that the clinical priorities are declared without using computer technicalities; they should be declared in the clinicians' language to create a comfortable environment among the workshop attendees. 
This workshop also provides an opportunity for stakeholders to discuss which priorities are critical, which priorities are not relevant, who are the clinicians in charge of the priorities, and other additional information. 
To conduct the workshop, the D\&I Framework recommends that stakeholders should be divided into groups. 
Additionally, in each group, an IT professional must lead the conversation. 
The mission of this IT professional is (i) to collect the priorities defined in each group and (ii) prevent clinicians from describing priorities not applicable in the clinical software\footnote{In our previous experience, some clinicians described priorities that could not be addressed within established deadlines}.
Finally, the workshop concludes with a discussion of all the priorities defined by the groups. 
At this point, the project leader can add, modify, or remove priorities in the D\&I Framework platform.

\subsection{Collaborative selection of implementation strategies {\large \textcircled{\small 3}}}

The objective of this stage is to select the implementation or dissemination strategies (more details on the strategies can be found in \cite{marquez2020}). 
This stage involves the same stakeholders as in the previous stage. 
For each group, stakeholders use one session of the D\&I Framework platform to determine which implementation or dissemination strategies are appropriate to address the clinical priorities. 
Each strategy is classified by category and color; each color represents a category. 
The platform also allows for drag-and-drop action for the ease of stakeholders in associating priorities with strategies. 
Since stakeholders can estimate that a priority can be addressed by one or several strategies, the platform offers the facility to assign a priority to several strategies.

\subsection{Analysis {\large \textcircled{\small 4}}}

\begin{figure} [h]
  \centering
  \includegraphics[scale=0.185]{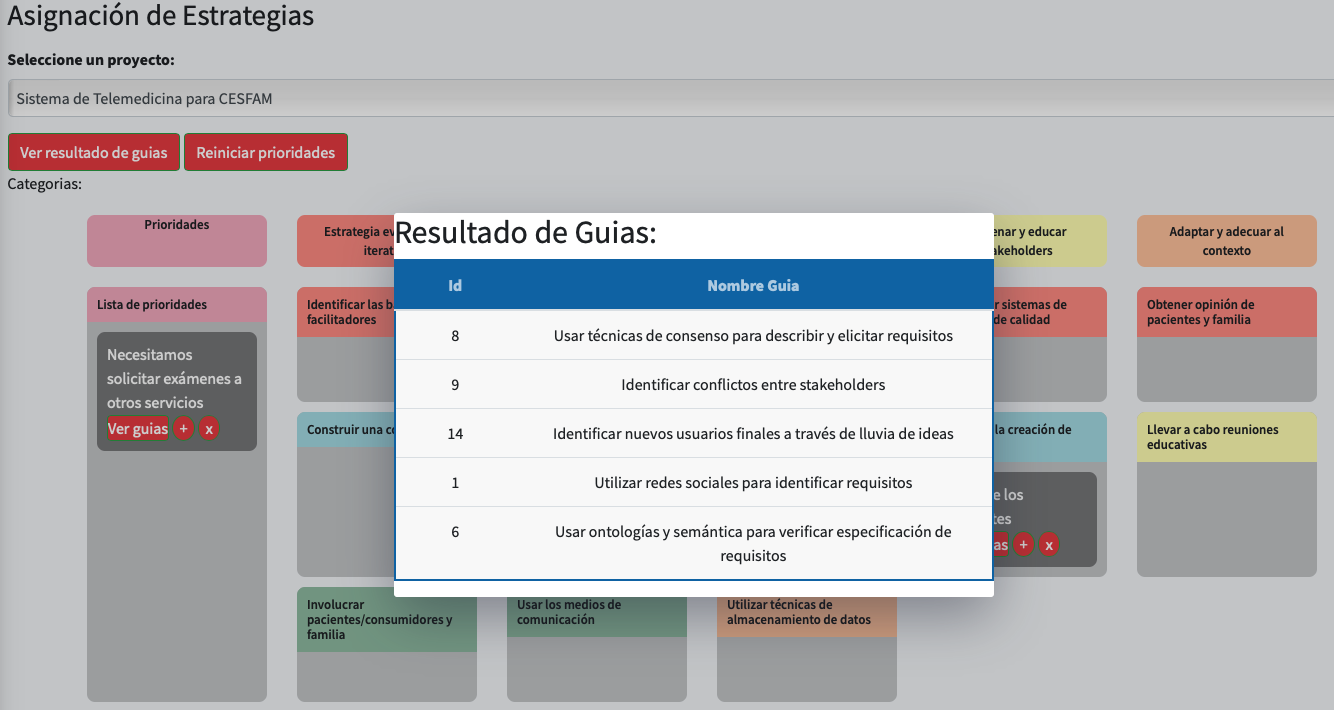}
  \caption{Example of guidelines to elicit requirements generated by the D\&I Framework}
  \label{fig:example}
\end{figure}

When the workshop concludes, the development team discusses which strategies were selected by the stakeholders. 
For each strategy selected, the D\&I Framework platform suggests a set of requirement elicitation guidelines to address such strategies.  
For example, suppose for a clinical priority, stakeholders have established that three implementation strategies are appropriate for intervening such priority. 
The D\&I Framework platform then shows which guidelines are appropriate to operationalize the priority intervention (see Figure \ref{fig:example}) based on the three strategies selected. 
In this example, the proposed guidelines are as follows:

\begin{itemize}
    \item Use consensus techniques to describe and elicit requirements (Id=8)
    
    \item Identify conflicts between stakeholders (Id=9)
    
    \item Identify new end-users through brainstorming (Id=14)
    
    \item Use social networks to identify requirements (Id=1)
    
    \item Use ontologies and semantics to verify requirements specification (Id=6)
    
\end{itemize}

These suggestions are based on a catalog described in \cite{marquez2020}, which maps implementation and dissemination strategies towards requirements elicitation techniques. 
Finally, the development team must select the appropriate guidelines for requirements elicitation and then proceed with the corresponding analysis and documentation.
\section{Case study}
\label{sec:caseStudy}

In this section, we describe the design of a case study where we evaluate the perception of clinicians regarding functionalities and tasks implemented in a clinical software. 
These functionalities and tasks were elicited and characterized through the D\&I Framework.

\subsection{Context}

Bed management corresponds to a strategic and support unit within clinical institutions that allows for the optimization of bed resources and the transfer of patients according to the care they require \cite{Boaden1999}. 
One of the most critical restrictions when treating patients or being able to respond to different treatments is the number and use of beds. 
These restrictions can severely compromise the coordination and management of beds in a clinical facility.

To support the management of clinical beds in the Chilean health network, we have developed an intelligent system for the management and analysis of the bed allocation called SIGICAM \cite{Taramasco2019} \cite{taramasco2019(2)}. 
This system allows systematizing people's hospitalization in health centers of low, medium or high complexity. 
Its purpose is to reduce waiting times in hospitals through intelligent and systematized administration of the bed allocation processes and patient transfer, using optimization techniques. 
These techniques allow for the detection of restrictions and critical variables that affect management and they serve as a support tool for decision making.

During the development of SIGICAM, the D\&I Framework supported the developers in the process of elicitation and description of requirements.
Indeed, the framework allowed the identification of the clinical problem using the perspective of all the clinicians who are related to the bed management process in the hospital.

\subsection{Goal}

From the release to production until 2020, SIGICAM has had three new releases that have improved the system functionalities.
In all those releases, the D\&I Framework has been used to elicit and describe the corresponding requirements. 
Although the D\&I Framework has supported the SIGICAM development team in each release, it is unclear if the end-users (clinicians) perceive such improvements in the system. 
Therefore, the study goal, established using the Goal-Question-Metric (GQM) approach \cite{Basili1992}, is to \textit{analyze} the functionalities and tasks of SIGICAM \textit{for the purpose of} evaluating the impact of using the D\&I Framework \textit{with respect to} the elicitation and description of software requirements \textit{from the point of view of} clinicians \textit{in the context of} bed management.

\subsection{Research questions}

We established the following research question:

\vspace{0.2cm}
\begin{centering}
\fbox{\begin{minipage}{24.5em}
\textit{Is there a difference in clinicians' perception about SIGICAM's functionalities and tasks regarding the first release (November 2018) and the last one (February 2020)?}
\end{minipage}}
\end{centering}
\vspace{0.2cm}

There are several sub-processes, health professionals and external entities (such as laboratories, providers, exams, among others) that the software development team must understand in the context of bed management. 
Despite the fact that Software Engineering has a plethora of techniques and processes to evaluate, analyze and validate software releases, there is little information about the perception of clinicians regarding whether SIGICAM effectively implements all the functionalities and tasks associated with bed management. 
More precisely, there is a lack of discussion about whether the requirements elicitation guidelines suggested by SIGICAM were useful to the development team in identifying and implementing those software requirements critical for the clinicians.

\subsection{Case and subject selection}

SIGICAM has several modules that allow bed management. 
For this case study, we used the four most critical modules of the system, which are as follows:

\begin{itemize}
    \item \textit{Hospital bed optimization module}: This module considers optimization algorithms to support the internal sub-processes of bed allocation and patient transfer.
    
    \item \textit{Synchronization module with the hospital's Central Bed Management Unit}: This module is a communication channel between the service that uses the beds and the hospital's central bed unit.
    
    \item \textit{Synchronization module with the Medical-Statistical Guidance Service}: This service manages the admission of elderly patients to the health network. For this reason, when an elderly patient is admitted to a hospital, a bed must be assigned to him/her.
    
    \item \textit{Reporting module}: This module uses SIGICAM knowledge to generate statistical reports for different purposes.
\end{itemize}

The subjects participating in this case study correspond to emergency nurses, floor nurses, bed managers and service directors.

\subsection{Data collection procedure}

In this study, we focused on evaluating end-users' perceptions based on the usability of SIGICAM. 
We have selected usability because this quality attribute is one of the most important for clinicians \cite{Peute2007}.

We used an instrument called the Health Information Technology Usability Evaluation Scale (Health-ITUES) \cite{Yen2010} to evaluate the usability of SIGICAM.
This instrument is a questionnaire based on four factors, which are as follows: quality of work-life ({\large \textcircled{\small A}}), perceived usefulness ({\large \textcircled{\small B}}), perceived ease of use ({\large \textcircled{\small C}}) and user control ({\large \textcircled{\small D}}). 
Each of these factors, in turn, is composed of questions measured on a 5-Likert scale (see Table \ref{tab:healthITUES}).
The goal of Health-ITUES is to measure the relationship between users and system tasks in a specified setting.

\begin{table}[h]
\caption{The Health-ITUES instrument}

\centering{}%
\begin{tabular}{|c|c|p{6.3cm}|}
\hline 
\cellcolor{gray!15}Factor & \cellcolor{gray!15}ID & \cellcolor{gray!15}\hspace{2.2cm} Questions\tabularnewline
\hline 
\multirow{3}{*}{{\large \textcircled{\small A}}} & A1 & \emph{I think SIGICAM has been a positive contribution to nursing
work}\tabularnewline
\cline{2-3} 
 & A2 & \emph{I think SIGICAM has been a positive contribution to the hospital}\tabularnewline
\cline{2-3} 
 & A3 & \emph{The technology delivered by SIGICAM is an essential part of
the bed management and analysis process}\tabularnewline
\hline 
\multirow{9}{*}{{\large \textcircled{\small B}}} & B4 & \emph{Using SIGICAM makes it easy to request available beds}\tabularnewline
\cline{2-3} 
 & B5 & \emph{SIGICAM makes it possible to manage beds more quickly}\tabularnewline
\cline{2-3} 
 & B6 & \emph{SIGICAM increases the probability of assigning or reassigning
a bed to a patient}\tabularnewline
\cline{2-3} 
 & B7 & \emph{SIGICAM is useful for requesting beds and managing patients
on waiting lists}\tabularnewline
\cline{2-3} 
 & B8 & \emph{I think SIGICAM presents a more equitable process for bed management}\tabularnewline
\cline{2-3} 
 & B9 & \emph{I am satisfied with SIGICAM for managing and analyzing the provision
of beds in the healthcare network}\tabularnewline
\cline{2-3} 
 & B10 & \emph{I can perform bed management tasks promptly due to the use of
SIGICAM}\tabularnewline
\cline{2-3} 
 & B11 & \emph{SIGICAM increases effectiveness in hospital waiting times}\tabularnewline
\cline{2-3} 
 & B12 & \emph{I am able to fulfill all my assigned tasks using SIGICAM}\tabularnewline
\hline 
\multirow{5}{*}{{\large \textcircled{\small C}}} & C13 & \emph{I'm comfortable with my ability to use SIGICAM}\tabularnewline
\cline{2-3} 
 & C14 & \emph{Learning to use SIGICAM is easy for me}\tabularnewline
\cline{2-3} 
 & C15 & \emph{It's easy for me to be proficient in the use of SIGICAM}\tabularnewline
\cline{2-3} 
 & C16 & \emph{SIGICAM is easy for me to use}\tabularnewline
\cline{2-3} 
 & C17 & \emph{I can always remember how to start and use SIGICAM}\tabularnewline
\hline 
\multirow{3}{*}{{\large \textcircled{\small D}}} & D18 & \emph{SIGICAM shows error messages that tell me clearly how to solve
problems}\tabularnewline
\cline{2-3} 
 & D19 & \emph{If I make mistakes in SIGICAM, I can solve it easily and quickly}\tabularnewline
\cline{2-3} 
 & D20 & \emph{The information (such as online help, on-screen messages and
other documentation) provided with SIGICAM is clear}\tabularnewline
\hline 
\end{tabular}
\label{tab:healthITUES}
\end{table}

\subsection{Analysis procedure}

We compared the responses to the Health-ITUES questionnaire carried out in November 2018 and February 2020. 
The first questionnaire was executed three months after the first release of SIGICAM into production; 50 clinicians participated in the questionnaire. 
On the other hand, the second questionnaire was executed three months after the third release; 48 clinicians participated in this second questionnaire.
To analyze the answers to both questionnaires, we used descriptive statistics.
We first calculate the average of the answers for each question of the questionnaire.
Then, we compared the averages of the answers of each question in both questionnaires.
Finally, we determine whether there is a difference between the questionnaires' averages using the $t$-test.

\subsection{Validity}

As suggested by Runeson \textit{et al.} \cite{Runeson2009}, we address threats to validity of our study in the following aspects: construct validity, internal validity, external validity and reliability.

\paragraph{Construct validity} This validity is concerned with the generalization of the results. 
One of the main threats to this validity is the worry of clinicians to evaluate SIGICAM. 
To mitigate this threat, before running the questionnaire, we inform respondents that the survey is anonymous and that the survey results are used for academic purposes. 
Another threat we have identified is the use of one type of measurement in this study.  
Although we are aware of this threat, we have decided to use Health-ITUES as an instrument because it has been used in different studies with significant results \cite{Yen2014}. 
Although we cannot generalize results using this instrument alone, we intend to use more instruments to evaluate SIGICAM as we move forward with our research. 
    
\paragraph{Internal validity} This validity is related to the influences that can affect the independent variables with respect to the study's causality. 
A detected threat in this validity aspect is the instrument used to conduct the case study. 
Although evaluating a software product involves considering different perspectives and techniques, we mitigate this threat by selecting a perspective and an instrument created to evaluate usability in health information technologies. 
Another detected threat is the intention of the subjects to imitate the responses of others. 
To mitigate this, we have established that the survey is individual and we mention that the survey is a personal evaluation of SIGICAM's functionalities.
    
\paragraph{External validity} This aspect of validity is concerned with how widely the results of the findings can be generalized. 
The main threat detected is to have a subject population that is not representative of the population we want to generalize. 
To mitigate this threat, we have selected a questionnaire oriented to clinicians. 
Therefore, the subjects used in the case study must have a clinical profile.
    
\paragraph{Relaibility} This validity refers to the extent to which the data and analysis are dependent on the specific researchers. 
The main threat to this validity points to the confidence in the measurements made in the study. 
To mitigate this threat, we have used widely known descriptive statistics methods.
Furthermore, we used a quality evaluation questionnaire tested in several studies.

\section{Results}
\label{sec:results}

\begin{figure*} [h]
  \centering
  \begin{tikzpicture}
\footnotesize
\begin{axis}[
    width=18.5cm, height=5.0cm,
    symbolic x coords={A1,A2,A3,B4,B5,B6,B7,B8,B9,B10,B11,B12,C13,C14,C15,C16,C17,D18,D19,D20},
    ytick={0,0.5,1,1.5,2,2.5,3,3.5,4,4.5,5},
    ybar,
    ylabel= Average Health-ITUES responses,
    xlabel= Health-ITUES questions,
    every node near coord/.append style={rotate=90, anchor=west},
	bar width=7pt,
	enlargelimits=0.05,
    legend style={at={(0.5,-0.35)},
      anchor=north,legend columns=-1},
    xtick=data,
    nodes near coords,
    nodes near coords align={vertical},
    axis x line*=bottom,
    axis y line*=left,
    ]

\addplot [fill=green!50] coordinates {(A1,3.1)
(A2,3.05)
(A3,3.35)
(B4,2.55)
(B5,2.45)
(B6,2.5)
(B7,3)
(B8,2.85)
(B9,2.55)
(B10,2.55)
(B11,2.5)
(B12,2.55)
(C13,2.95)
(C14,3.45)
(C15,3.1)
(C16,3.4)
(C17,3.9)
(D18,2.7)
(D19,2.8)
(D20,2.9)};

\addplot [fill=blue!50] coordinates {
(A1,3.45833333333333)
(A2,3.47916666666667)
(A3,3.58333333333333)
(B4,2.6875)
(B5,2.64583333333333)
(B6,3.16666666666667)
(B7,3.0625)
(B8,3.22916666666667)
(B9,3.14583333333333)
(B10,3.10416666666667)
(B11,2.83333333333333)
(B12,2.70833333333333)
(C13,3.60416666666667)
(C14,3.66666666666667)
(C15,3.72916666666667)
(C16,3.70833333333333)
(C17,4.4375)
(D18,2.91)
(D19,3.08333333333333)
(D20,2.89583333333333)};
\legend{\strut 2018,\strut 2020}
\end{axis}
\end{tikzpicture}
  \caption{Comparison of the average value of the answers to each Health-ITUES question for the years 2018 and 2020}
  \label{fig:distribution}
\end{figure*}
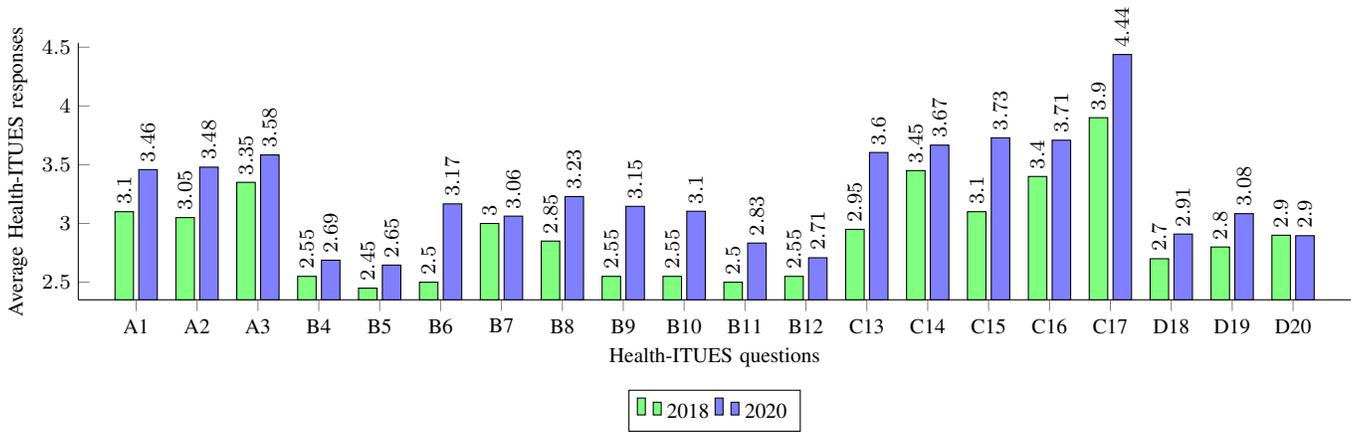

Figure \ref{fig:distribution} compares the average of the answers to each question of the Health-ITUES questionnaire carried out in 2018 and 2020. 
For instance, in question A1, the average in 2018 was 3.1 and in 2020, 3.46.
Additionally, Table \ref{tab:statsResults} describes (i) the global average, (ii) the average of the responses for each Health-ITUES factor, (iii) the median and (iv) the standard deviation (SD) of the averages per year. 
Before evaluating the statistical difference of the two questionnaires' answers, we tested the normality of each group of averages using the Shapiro-Wilk test \cite{Hanusz2016}. 
Once normality has been checked, we use the $t$-test to determine if there is a significant difference between the two groups of averages. 
We establish the following null hypothesis ($H_0$): \textit{there is no difference between the groups of averages of both questionnaires}. 
The results of the $t$-test show that the $t$-value is -2.60178; then, the $p$-value is 0.013147. 
These results allow us to refute $H_0$ ($p$ $<$ 0.05), i.e., the difference between both groups of averages is significant.

\begin{table}[h]

\caption{Global average ($A$), average of each category, mean and standard deviation ($SD$) of the results of the Health-ITUES questionnaires of the years 2018 and 2020.}

\centering{}%
\begin{tabular}{|c|c|c|c|c|c|c|c|}
\hline 
\cellcolor{gray!15}Year & \cellcolor{gray!15}$A$ & \cellcolor{gray!15}$A$ { \textcircled{\scriptsize A}} & \cellcolor{gray!15}$A$ {\textcircled{\scriptsize B}} & \cellcolor{gray!15}$A$ {\textcircled{\scriptsize C}} & \cellcolor{gray!15}$A$ {\textcircled{\scriptsize D}} & \cellcolor{gray!15}Media & \cellcolor{gray!15}$SD$\tabularnewline
\hline 
2018 & 2.91 & 3.17 & 2.61 & 3.36 & 2.80 & 2.87 & 0.39\tabularnewline
\hline 
2020 & 3.25 & 3.51 & 2.95 & 3.83 & 2.96 & 3.16 & 0.44\tabularnewline
\hline
\end{tabular}
\label{tab:statsResults}
\end{table}

\section{Discussion}
\label{sec:discussion}

The study results show a better perception of clinicians regarding the functionalities and tasks implemented in SIGICAM compared with 2018 and 2020.
According to the results described in Table \ref{tab:statsResults}, the overall average of the survey carried out in 2018 was 2.91 and the survey carried out in 2020 was 3.25. 
In addition, we realized that there is a significant difference in the responses of both questionnaires.
Although these preliminary results do not allow to bridge the requirements elicitation using the D\&I Framework with the results of the Health-ITUES questionnaire, we can interpret the results of the questionnaires from the point of view of the SIGICAM developers and the feedback of the clinicians. 

With regard to the first release of SIGICAM, the developers appreciated the guidelines proposed by the D\&I Framework to elicit requirements. 
They realized that using well-known elicitation techniques is not enough to understand the bed management process.
Requirements elicitation guidelines such as ``modeling the domain of the clinical service'', ``obtaining requirements from the bed management business model'', and ``identifying requirements through the characterization of the clinical professional-patient relationship'' were critical to identifying more pragmatic functional and non-functional requirements. 
Consequently, the first release of SIGICAM was well received by the hospital community\footnote{Press Note about the release of SIGICAM. Published in December 2018 (in Spanish): \url{https://www.elmostrador.cl/agenda-pais/2018/12/30/con-inteligencia-artificial-se-organiza-la-reserva-de-camas-hospitalarias/}}. 

Subsequently, in all the improvements implemented in SIGICAM, the D\&I Framework was used to identify new requirements based on the clinical priorities concerning the hospital bed management process. 
Given that bed management is a highly changing process, in each release, the D\&I Framework allowed the developers to be aware of these changes in order to identify, characterize and implement them in the system.

Feedback from SIGICAM's developers and the study results also reveal that the implementation and dissemination strategies help to intervene in challenging and rapidly changing contexts. 
The D\&I Framework supported the developers in understanding how well an intervention (implementation or dissemination strategy) manages to have a positive effect on clinicians. 
The framework also helped to detect whether there are unintended consequences to implementing some strategies (and hence to the elicitation of requirements), as well as the cost, the possibility of generalization, and the potential for sustainability of the strategies. 

On the other hand, the Health-ITUES questionnaire also reveals interesting findings:

\paragraph{Quality of work-life ({\large \textcircled{\small A}})} This factor characterizes the impact of the system beyond the system's functionalities. For 2018, the average of each question (A1, A2 and A3) is above the overall average of the survey (2.91). This reveals to us that SIGICAM was able to improve, to some extent, the daily work done by clinicians with regard to bed management. In 2020 it was no exception. The results of the questionnaire were better than in 2018. This allows us to infer that the improvements implemented in SIGICAM's releases encourage clinicians to appreciate the system as a support to their daily work. 
    
\paragraph{Perceived Usefulness ({\large \textcircled{\small B}})} This factor evaluates the usefulness of a system for a specific task. Although there is an improvement in the averages related to this factor (B4 to B12) between 2018 and 2020, some specific SIGICAM tasks still need to be improved. The D\&I Framework has supported the development team in improving SIGICAM; however, other factors (such as low technology adoption) compromise some end-users' use of SIGICAM. Not all clinicians have the skills to use systems to support their daily work.
    
\paragraph{Perceived Ease of Use ({\large \textcircled{\small C}})} This factor points to the user-system interaction (C13 to C17). The results obtained in 2018 and 2020 describe that clinicians appreciate the ease of use of the system. The SIGICAM development team appreciates this because one of the most suggested guidelines of the D\&I Framework in the releases is to ``identify requirements through brainstorming and education sessions''. This guideline aims to bring together different end-users to identify their concerns and expectations of the system.
    
\paragraph{User Control ({\large \textcircled{\small D}})} This factor points to the user control ability (D18, D19 and D20). In questions D18 and D19, there was an improvement in the results of the questionnaire executed in 2020. Nevertheless, the results at D20 remained the same in both years. In spite of the above, some aspects of SIGICAM need further improvement. For instance, the capacity of the system's user control in specific tasks must be improved.

The Health-ITUES questionnaire results have revealed that, although the results obtained in 2020 are promising and describe that the system has progressed, some usability aspects need to be improved.  
On the other hand, SIGICAM developers and clinicians have valued the requirement elicitation strategies suggested by the D\&I Framework. 
Consequently, in each release, software developers and clinicians are increasingly reducing the communication gap between them. 
This implies that developers are more aware of the bed management process and understand clinicians' real needs. 
\section{Conclusions}
\label{sec:conclusions}

We have presented a study that evaluates the functionalities and task of a bed management system called SIGICAM from the point of view of clinicians. 
The identification and description of these functionalities and task were conducted through the D\&I Framework. 
This collaborative technique identifies clinical priorities to be mapped into implementation strategies and then into requirements elicitation guidelines.

We evaluated SIGICAM from the point of view of usability using the Health-ITUES questionnaire, which was carried out in 2018 and 2020. 
The results of the study indicate that clinicians perceive an improvement in the system. 
Although the Health-ITUES questionnaire is not related to the software requirements identified by the D\&I Framework, we can conclude that the framework's guidelines allowed SIGICAM developers to identify, characterize, and describe more pragmatic requirements that represent the bed management process. 
Clinicians also appreciate this more realistic representation of software requirements because the functionalities and tasks implemented in SIGICAM have favored and helped their daily work.

To further our research, we plan to evaluate whether the D\&I Framework has supported SIGICAM's developers in managing the COVID-19 contingency for bed management. 
Currently, SIGICAM is being used as a support system in two public hospitals in Chile. 
Due to the pandemics' current scenario, the system has had to adapt to all the changes related to the management of COVID-19 positive patients and bed allocation. 
Therefore, we are planning to conduct a study to evaluate the perception of clinicians in both hospitals regarding the use of SIGICAM and the COVID-19 contingency.
\section*{Acknowledgments}
\label{sec:ack}

This research is supported by the project ``Sistema inteligente para la gestión y análisis de la dotación de camas en la red asistencial del sector público'' FONDEF - Ideas (ID16I10449). We also thank the National Center for Health Information Systems (\textit{Centro Nacional en Sistemas de Información en Salud, CENS)}, Chile for the contribution in this study

\bibliographystyle{ieeetr}
\bibliography{biblio}

\begin{thebibliography}{10}

\bibitem{bourque2014}
P.~Bourque and R.~E. Fairley, ``Guide to the software engineering body of
  knowledge ({SWEBOK} ({R})): Version 3.0.,'' {\em IEEE Computer Society
  Press}, 2014.

\bibitem{Van2009}
A.~Van~Lamsweerde, {\em Requirements engineering: From system goals to {UML}
  models to software}.
\newblock John Wiley \& Sons, 2009.

\bibitem{kaplan2009}
B.~Kaplan and K.~D. Harris-Salamone, ``Health it success and failure:
  recommendations from literature and an {AMIA} workshop,'' {\em Journal of the
  American Medical Informatics Association}, vol.~16, no.~3, pp.~291--299,
  2009.
\newblock \mbox{doi}:\hspace{0.1cm}\url{{
  https://doi.org/10.1197/jamia.M2997}}.

\bibitem{heeks2006}
R.~Heeks, ``Health information systems: Failure, success and improvisation,''
  {\em International journal of medical informatics}, vol.~75, no.~2,
  pp.~125--137, 2006.
\newblock \mbox{doi}:\hspace{0.1cm}\url{{
  https://doi.org/10.1016/j.ijmedinf.2005.07.024}}.

\bibitem{marquez2020}
G.~Márquez and C.~Taramasco, ``Using dissemination and implementation
  strategies to evaluate requirement elicitation guidelines: A case study in a
  bed management system,'' {\em IEEE Access}, vol.~8, pp.~145787--145802, 2020.
\newblock \mbox{doi}:\url{{ 10.1109/ACCESS.2020.3015144}}.

\bibitem{brownson2017}
R.~C. Brownson, G.~A. Colditz, and E.~K. Proctor, {\em Dissemination and
  implementation research in health: translating science to practice}.
\newblock Second Edition. Oxford University Press, 2017.

\bibitem{Yen2010}
P.~Y. Yen, D.~Wantland, and S.~Bakken, ``Development of a customizable health
  {IT} usability evaluation scale,'' {\em {AMIA} Annual Symposium Proceedings},
  vol.~2010, p.~917–921, 2010.

\bibitem{Strasser2011}
M.~Strasser, F.~Pfeifer, E.~Helm, A.~Schuler, and J.~Altmann, ``Defining and
  reconstructing clinical processes based on {IHE} and {BPMN} 2.0,'' {\em
  Studies in health technology and informatics}, vol.~169, pp.~482--486, 2011.

\bibitem{Islam2016}
R.~Islam, C.~Weir, and G.~Del~Fiol, ``Clinical complexity in medicine: a
  measurement model of task and patient complexity,'' {\em Methods of
  Information in Medicine}, vol.~55, no.~1, p.~14, 2016.
\newblock \mbox{doi}:\url{{ 10.3414/ME15-01-0031}}.

\bibitem{tiwari2012}
S.~Tiwari, S.~S. Rathore, and A.~Gupta, ``Selecting requirement elicitation
  techniques for software projects,'' {\em {CSI} Sixth International Conference
  on Software Engineering ({CONSEG})}, pp.~1--10, 2012.
\newblock \mbox{doi}:\url{{ 10.1109/CONSEG.2012.6349486}}.

\bibitem{charette2005}
R.~N. Charette, ``Why software fails [software failure],'' {\em IEEE spectrum},
  vol.~42, no.~9, pp.~42--49, 2005.
\newblock \mbox{doi}:\hspace{0.1cm}\url{{ 10.1109/MSPEC.2005.1502528}}.

\bibitem{doloi2013}
H.~Doloi, ``Cost overruns and failure in project management: Understanding the
  roles of key stakeholders in construction projects,'' {\em Journal of
  construction engineering and management}, vol.~139, no.~3, pp.~267--279,
  2013.
\newblock \mbox{doi}:\hspace{0.1cm}\url{{
  https://doi.org/10.1061/(ASCE)CO.1943-7862.0000621}}.

\bibitem{cabana1999}
M.~D. Cabana, C.~S. Rand, N.~R. Powe, A.~W. Wu, M.~H. Wilson, P.~A.~C. Abboud,
  and H.~R. Rubin, ``Why don't physicians follow clinical practice guidelines?:
  A framework for improvement,'' {\em Jama}, vol.~282, no.~15, pp.~1458--1465,
  1999.
\newblock \mbox{doi}:\hspace{0.1cm}\url{{ 10.1001/jama.282.15.1458}}.

\bibitem{johnson2011}
C.~W. Johnson, ``Identifying common problems in the acquisition and deployment
  of large-scale, safety–critical, software projects in the {US} and {UK}
  healthcare systems,'' {\em Safety Science}, vol.~49, no.~5, pp.~735--745,
  2011.
\newblock \mbox{doi}:\hspace{0.1cm}\url{{
  https://doi.org/10.1016/j.ssci.2010.12.003}}.

\bibitem{chowdhury2007}
R.~Chowdhury, R.~E. Butler, and S.~Clarke, ``Healthcare {IT} project failure: A
  systems perspective,'' {\em Journal of Cases on Information Technology
  ({JCIT})}, vol.~9, no.~1, pp.~1--15, 2007.
\newblock \mbox{doi}:\hspace{0.1cm}\url{{10.4018/jcit.2007100101}}.

\bibitem{johnson2006}
C.~W. Johnson, ``Why did that happen? exploring the proliferation of barely
  usable software in healthcare systems,'' {\em BMJ Quality \& Safety},
  vol.~15, no.~i76-i81, p.~suppl 1, 2006.
\newblock \mbox{doi}:\hspace{0.1cm}\url{{ 10.1136/qshc.2005.016105}}.

\bibitem{Vamos2020}
C.~A. Vamos, S.~M. Green, S.~Griner, E.~Daley, R.~DeBate, T.~Jacobs, and
  S.~Christiansen, ``Identifying implementation science characteristics for a
  prenatal oral health ehealth application,'' {\em Health Promotion Practice},
  vol.~21, no.~2, pp.~246--258, 2020.
\newblock \mbox{doi}:\url{{ https://doi.org/10.1177/1524839918793628}}.

\bibitem{Richardson2012}
J.~E. Richardson, E.~L. Abramson, E.~R. Pfoh, R.~Kaushal, and H.~Investigators,
  ``Bridging informatics and implementation science: evaluating a framework to
  assess electronic health record implementations in community settings,'' {\em
  AMIA Symposium}, p.~770–778, 2012.

\bibitem{Glasgow2014}
R.~E. Glasgow, S.~M. Phillips, and M.~A. Sanchez, ``Implementation science
  approaches for integrating ehealth research into practice and policy,'' {\em
  International journal of medical informatics}, vol.~83, no.~7, pp.~e1--e11,
  2014.
\newblock \mbox{doi}:\url{{ }}.

\bibitem{Schoville2015}
R.~R. Schoville and M.~G. Titler, ``Guiding healthcare technology
  implementation: a new integrated technology implementation model,'' {\em CIN:
  Computers, Informatics, Nursing}, vol.~33, no.~3, pp.~99--107, 2015.
\newblock \mbox{doi}:\url{{ 10.1097/CIN.0000000000000130}}.

\bibitem{Dykes2020}
P.~C. Dykes, G.~Lowenthal, A.~Faris, R.~Hack, D.~Harding, A.~Hurley, and P.~An,
  ``An implementation science approach to promote optimal implementation,
  adoption, use, and spread of continuous clinical monitoring system
  technology,'' {\em Journal of Patient Safety}, vol.~17, no.~1, pp.~56--62,
  2020.
\newblock \mbox{doi}:\url{{ 10.1097/pts.0000000000000790}}.

\bibitem{Boaden1999}
R.~Boaden, N.~Proudlove, and M.~Wilson, ``An exploratory study of bed
  management,'' {\em Journal of management in medicine}, vol.~13, no.~4,
  pp.~234--250, 1999.
\newblock \mbox{doi}:\url{{ https://doi.org/10.1108/02689239910292945}}.

\bibitem{Taramasco2019}
C.~Taramasco and R.~Olivares, ``{SIGICAM}: A new software to improve the
  patient care supported by a constraint-based model,'' {\em Studies in health
  technology and informatics}, vol.~264, pp.~824--828, 2019.
\newblock \mbox{doi}:\url{{ 10.3233/shti190338}}.

\bibitem{taramasco2019(2)}
C.~Taramasco, R.~Olivares, R.~Munoz, R.~Soto, M.~Villar, and V.~H.~C.
  de~Albuquerque, ``The patient bed assignment problem solved by autonomous bat
  algorithm,'' {\em Applied Soft Computing}, vol.~81, p.~105484, 2019.
\newblock \mbox{doi}:\hspace{0.1cm}\url{{
  https://doi.org/10.1016/j.asoc.2019.105484}}.

\bibitem{Basili1992}
V.~R. Basili, ``Software modeling and measurement: the goal/question/metric
  paradigm,'' {\em Technical Report, University of Maryland at College Park},
  1992.

\bibitem{Peute2007}
L.~W. Peute and M.~W. Jaspers, ``The significance of a usability evaluation of
  an emerging laboratory order entry system,'' {\em International journal of
  medical informatics}, vol.~76, no.~2-3, pp.~157--168, 2007.
\newblock \mbox{doi}:\url{{ https://doi.org/10.1016/j.ijmedinf.2006.06.003}}.

\bibitem{Runeson2009}
P.~Runeson and M.~Höst, ``Guidelines for conducting and reporting case study
  research in software engineering,'' {\em Empirical software engineering},
  vol.~14, no.~2, p.~131, 2009.
\newblock \mbox{doi}:\url{{ https://doi.org/10.1007/s10664-008-9102-8}}.

\bibitem{Yen2014}
P.~Y. Yen, K.~H. Sousa, and S.~Bakken, ``Examining construct and predictive
  validity of the health-it usability evaluation scale: confirmatory factor
  analysis and structural equation modeling results,'' {\em Journal of the
  American Medical Informatics Association}, vol.~21, no.~e2, pp.~e241--e248,
  2014.
\newblock \mbox{doi}:\url{{ https://doi.org/10.1136/amiajnl-2013-001811}}.

\bibitem{Hanusz2016}
Z.~Hanusz, J.~Tarasinska, and W.~Zielinski, ``Shapiro-wilk test with known
  mean,'' {\em {REVSTAT}-Statistical Journal}, vol.~14, no.~1, pp.~89--100,
  2016.

\end{thebibliography}

\end{document}